# A Feasibility study for Deep learning based automated brain tumor segmentation using Magnetic Resonance Images

Shanaka Ramesh Gunasekara[1], HNTK Kaldera[2], Maheshi B. Dissanayake[3]
Dept. Of Electrical and Electronic Engineering, Faculty of Engineering,
University of Peradeniya 20400, Sri Lanka.
[1] shanakag@eng.pdn.ac.lk , [2] nikalal.k@eng.pdn.ac.lk, [3] maheshid@eng.pdn.ac.lk

**Abstract**—Deep learning algorithms have accounted for the rapid acceleration of research in artificial intelligence in medical image analysis, interpretation, and segmentation with many potential applications across various sub disciplines in medicine. However, only limited number of research which investigates these application scenarios, are deployed into the clinical sector for the evaluation of the real requirement and the practical challenges of the model deployment. In this research, a deep convolutional neural network (CNN) based classification network and Faster RCNN based localization network were developed for brain tumor MR image classification and tumor localization. A typical edge detection algorithm called Prewitt was used for tumor segmentation task, based on the output of the tumor localization. Overall performance of the proposed tumor segmentation architecture, was analyzed using objective quality parameters including Accuracy, Boundary Displacement Error (BDE), Dice score and confidence interval. A subjective quality assessment of the model was conducted based on the Double Stimulus Impairment Scale (DSIS) protocol using the input of medical expertise. It was observed that the confidence level of our segmented output was in a similar range to that of experts. Also, the Neurologists have rated the output of our model as highly accurate segmentation.

—————————◆—————————

## 1 INTRODUCTION

Magnetic resonance imaging (MRI) is one of the most common types of imaging technique in Radiology which is used for diagnosing blood vessel damages, brain injuries and tumors, strokes, eye problems, bone infections, heart disease, and many more, which would otherwise require invasive procedures. MRIs are widely used for Brain tumor diagnosis. The Brain tumor is an abnormal growth in brain cells that leads to considerable morbidity or mortality. There are more than 120 tumor types and all of those belong to two main tumor groups; primary brain tumors and secondary brain tumors. Typically, at the initial stage of diagnosis of tumors, non-invasive medical imaging techniques such as MRI and Computer Tomography (CT) are widely used by the physicians[1].

During the past 10 years, with the rapid improvement of artificial intelligence (AI) and deep learning algorithm, image classification and segmentation architectures and their applications have exhibited incredible performance growth. The improvements in these technologies directly affect the medical imaging sector[2]. Many researchers focused on developing more advance, accurate algorithms that are useful and powerful in medical image analysis. As a result, computerized systems were able to interpret the medical imaging outputs such as MRI or CT scans with high efficiency and accuracy, while getting closer to the diagnosis of a medical practitioner. Therefore, AI-radiologists can provide substantial benefits in many clinical settings, such as to prioritize workflow, support clinical decision making, large-scale screening initiatives and many more[3][4].

In the recent past, there was a rapid acceleration in academic research in the health care sector investigating the capabilities of the AI, specifically Deep learning, in medical image diagnosis. All of these algorithms were developed to match the performance of medical experts in a wide range of image analysis is tasks including, skin cancer diagnosis and classification, retinopathy detection for diabetics[5], automated diagnosis from chest imaging for lung module detection[6], pulmonary tuberculosis classification[7] and chest radiography diagnosis[8][9], breast cancer detection and diagnosis with mammography[10][11]. But only a few of these researches, present feedback of the design from the medical experts' point of view and feasibility study of the implementation of these developed algorithms in real world applications. For instance [12], explore the main limitations and challenges of the AI models developed for medical sector when translating them into clinical practice.

The majority of the machine learning studies have been retrospective; in both training and testing they used historically labeled data. Therefore when it comes to the real world scenario, the performance of the AI systems might be worse as the real world data differ from the training data set [12].Hence, it is necessary to carry out prospective studies with the real field data before the implementation. Note that, although there are many objective quality metrics that implies the performance of the developed AI algorithms, none of these reflect the clinical applicability of the algorithm. Researchers need to emphasize how the proposed AI and deep learning models may improve patient care with relatable facts[13]. One potential approach is to compare the results in a subjective manner with medical experts' opinions and obtain the subjective quality analysis of the algorithm with respect to the clinical needs.

In this research article, we present a subjective quality analysis of a proposed deep learning algorithm for brain tumor classification and segmentation. The section 2 summarizes the theoretical background and overview of the existing works in the literature related to the design problem. In section 3 we present the methodology embraced with data preparation, tumor classification and localization, and finally segmentation. The results and findings of the subjective performance evaluation are presented and discussed in section 4. Finally, section 5 concludes the paper.

## 2 BACKGROUND WORK

### 2.1 Initiation towards CNN

A network architecture which consists of convolutional, activation, max pooling and fully connected layers is defined as a Convolutional neural network (CNN). CNNs are powerful tools used in a wide range of applications, such as classification of images, segmentation of images, etc. In CNN, a filter or kernal slides over the input image and generates a feature map. Through different sizes of kernal filters, the network can be customized to generate numerous feature maps and increase the ability to learn more correlation between neighboring pixels of the image. But increasing the number of filters may result in huge computational cost as well as taking a long time to train the model. Furthermore, it can be more effective to look at smaller portions of the image for the interesting objects, rather than looking at an entire image at once to find certain features.

A CNN architecture blended with a region proposal network (RPN) is called Faster R-CNN. Basically, R-CNN is a classifier that classifies regions in an image as belonging to the object classes or not, while RPN creates regions of interest, called as anchors, for the input image and then predicts the probability of an anchor being in the background or foreground[14][15]. Generally, there are 9 standard anchors for the RPN network, with different dimensions' as illustrated in Fig01. In the RPN, it examines every anchor box to predict whether it contains an object or not. Furthermore, if there exists more than one object in the input image, RPN will output several anchor boxes with the relevant locations of each object. Later, RPN uses Region of Interest (ROI) pooling to identify the correct features belongs to a particular object. Afterward, R-CNN uses the proposed anchors from the RPN and classifies the object inside the bounding box to the existing classes. Also in this stage, the network fine tune the bounding box coordinates to fit in the ROI.

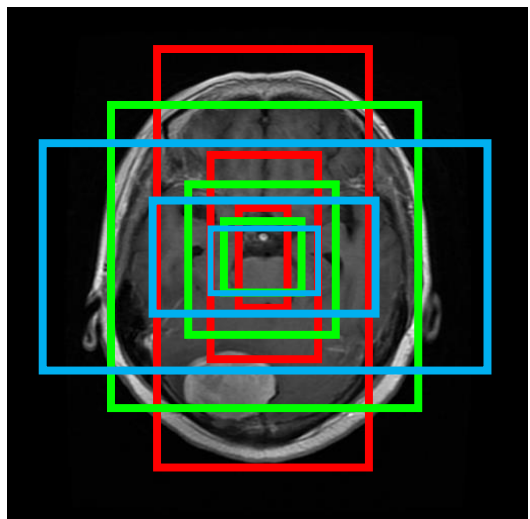

Fig 1 Anchors used by RPN on input image

## 2.2 Subjective quality analysis for medical images

Medical image analysis is only a tool for efficient diagnosis. The image analysis and the output should meet the requirement and decision criteria of the physician, i.e the experts. Although most of these objective quality analysis metrics are developed based on properties of the human vision system (HVS), it is impossible to model all the possible image distortions and levels which can identify with HVS along with human perception[16]. Hence, objective quality measures, would not justify the accuracy of the model, despite the promising numbers, specifically in the field of medical imaging. To overcome the diagnostic deficiencies of objective quality assessment in medical imaging, subjective quality assessment, where experts in the medical field give their best judgment on the perceptual quality and the diagnostic information of the output, is widely recommended.

International Telecommunications Union Recommendations (ITU-R) define subjective evaluation methods into three categories, Single Stimulus Impairment Scale(SSIS), Double Stimulus Impairment Scale (DSIS) and Double Stimulus Continuous Quality Scale (DSCQS)[16].As DSIS method is more suitable for assessing clearly visible impairments, we select DSIS procedure for our quality assessment. In DSIS method, five levels for impairment grading are defined and the reference image is shown with the processed image simultaneously. 5- Degree scale adopted in our study is defined as in Table (1).

Table 1: The DSIS Five Degree Scale

| Quality | Impairment |
|---|---|
| 5 Excellent | 5 Accurately Localized and segmented |
| 4 Good | 4 Perceptible, but not erroneous |
| 3 Fair | 3 Slightly erroneous |
| 2 Poor | 2 Erroneous |
| 1 Bad | 1 Very fallacious |

## 3 METHODOLOGY

### 3.1 Data preparation and preprocessing

T1 weighted MR images with 2D slices which were recorded at Guangzhou, China, and General Hospital, Tianjing Medical University, China, from 2005 to 2010 were used to develop and test the algorithm presented in this research[17]. The dataset consists of 3064 T1- weighted contrast-enhanced grayscale MR images, belonging to 233 different patients. Those images belonged to three tumor categories namely Meningioma (708 slices), Gliomas (1426 slices), and Pituitary (930 slices). In the research presented only axial MRIs of Meningioma and Glioma tumors were considered.

The implemented end to end system architecture was shown in Fig 02. The initial MRIs were of 512x512 resolution and at the preprocessing stage those images were down sampled to 128x128 to reduce the training time and the complexity of the network.

### 3.2 Network architecture for tumor classification and localization

A classical shallow convolutional neural network was developed to classify Meningioma and Glioma brain tumors by authors[18][19]. The CNN network consists of two convolutional layers, and each convolution layer is followed bya relu activation layer and a 2x2 maxpool layer. The first convolution layer has 3x3x20 kernal structure and it generates 20 different feature maps for every MR image input. The output of this convolution layer goes through a relu activation function and then 2x2 maxpooling layer. At the second convolution layer, there are 10 filters with the size of 3x3 and the outputs of the previous convolution layer are directly fed to the second set of Convolution, relu and maxpooling layers. The maxpooling layers reduce the dimensions of the output tensor by considering the max and min values of the processed window, which assists to avoid the model overfitting. After getting the 2D output feature map from the two convolution layers, it was converted into 1D feature map using a flattering operation and fed into a fully connected network for the classification task. The last layer of this fully connected network outputs the decision of the classification task either as Meningioma tumor or as Glioma tumor.

In the proposed model a faster R-CNN network follows the CNN network. The objective of the faster R-CNN is the tumor localization process. The RPN based faster R-CNN network is modeled following a transfer learning approach with the pre-trained weights generated using the COCO dataset, to detect the tumor boundaries. The classified output from the CNN is fed into the Faster R-CNN network and it generates estimated coordinates of the tumor bounding box as the output. As, these output coordinates of the bounding box belong to low resolution 128x128 MR image, it was mapped to the original 512x512 resolution at the post processing stage. Finally Prewitt contour algorithm is applied to the bounding box area for a finer tumor segmentation[20].

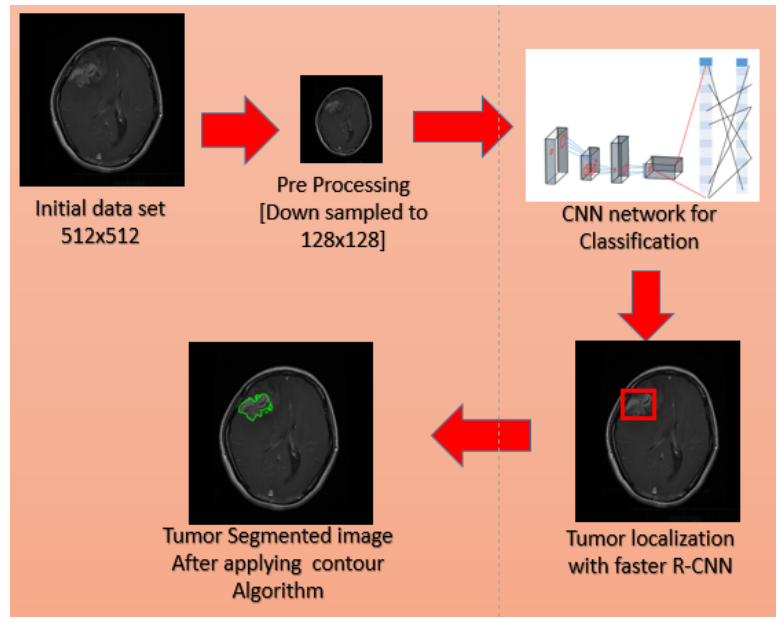

Fig 2: Architecture of the network

**3.3 Subjective quality evolution**

The main objective of the research presented is to conduct a feasibility study, where the output of the AI model presented is compared against the decision of the medical practitioners. With this objective we have carried out subjective quality evaluation of the output of the proposed tumor segmentation architecture.

As expressed in the ITU recommendations, 5 to 35number of subjects were required to perform a subjective equality assessment test and all of these subjects should have normal vision and be experts in the medical sector. In general15 number of subjects is sufficient to perform a good subjective quality evaluation. In our study, we have selected15 subjects, covering Neurologists, Medical Officers (MO)and Intern Home Officers (HO). None of these experts were given access to any details of the patient information or knowledge of the disease prevalence in the dataset used for this research. Before carrying out the assessment, a set of instructions consisting of the type of evaluation approach, the ranking scale, duration of the test was given to each subject. Environmental specifications were set to meet the room conditions, ambient lighting conditions, and viewing distance. Our subjective test was carried out by physically meeting selected evaluators with at least a good bachelor's degree in medicine, individually at their medical offices. About 10-15 minutes were given to doctors to read the instructions, adapt to the test environment and also lighting conditions. After that, a graphical user interface designed with Google Form was used for the evaluation and this user interface was designed in a simple manner to assist the evaluators with low computer skills. The sample of the graphical user interface used for the evaluation process is shown in Fig 03 and the link to the online questioner is at[21]. As in the Fig 03, the original MRI image and Segmented image are shown together according to ITU recommendations and the participants are asked to rate the accuracy of the prediction into 5 levels as well as to rate the accuracy of the prediction out of 100% at their discretion. 12 images each were used for Glioma and Meningioma for the evaluation process and all the data were recorded and analyzed. To make an unbiased evaluation process, MR images with incorrectsegmentations were added to the subjective test in both Glioma and Meningioma data sets.

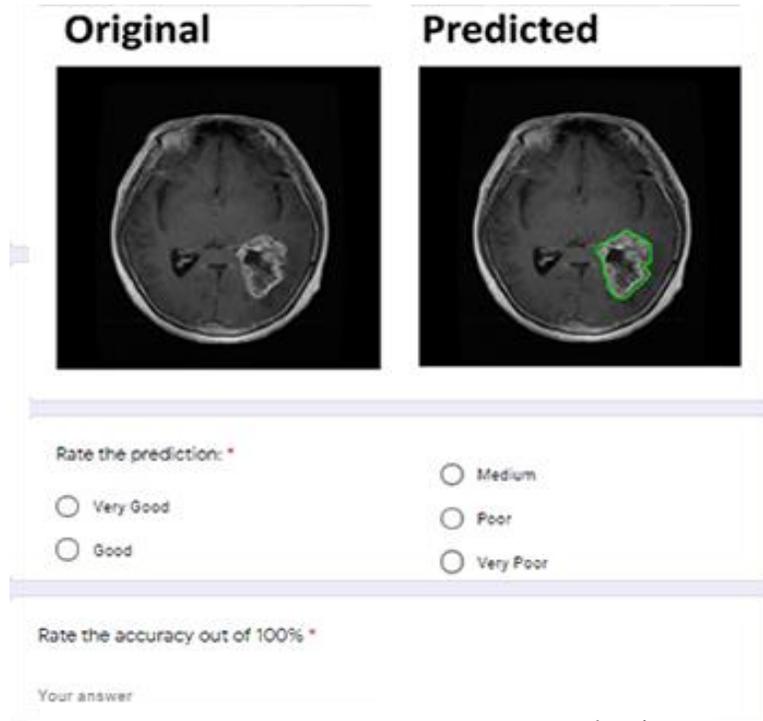

Fig 3 Graphical User Interface for DISS test evaluation

## 4 RESULTS AND DISCUSSION

The performance analysis was carried out on a lap top computer with Intel(R) 8 Series Chip set processor,4.00GB RAM and 64-bit operating system, ×64-basedprocessor. The objective quality evaluation of the segmentation is carried out using confidence interval, accuracy, Dice Score, and Boundary Displacement Error (BDE) metrics by comparing the tumor segmentation boundary of the proposed architecture with the ground truth. The results are tabulated in Table 02.

The objective parameters were calculated using standards definitions of them. Dice score is defined as the harmonic mean of precision and recall as in the eq. (1).

$$Dice\ Score = 2TP/(2TP + FP + FN) \quad (1)$$

The accuracy of each segmentation was calculated according to the eq. (2).

$$Accuracy = (TP + TN)/((TP + FP + TN + FN)) \quad (2)$$

The boundary displacement error, measures the average displacement error of one boundary pixels and the closet boundary pixels by means of the eq. (3)

$$\mu_{LA}(u,v) = \{\ (u-v)/(L-1)\ ;\ 0 < u - v \quad (3)$$

Here, TP is true positive, FP is False positive, FN is False negative and TN is True negatives.

Table 2: Overall performance of the system

| Boundary estimation method | Performance value |
|---|---|
| Confidence interval | 94.78 % |
| Dice score | 0.90 |
| Accuracy | 0.9125 |
| BDE | 3.581 |

As of our subject sample, neurologists are the most experienced and qualified personals on tumor detection using MRI. Hence, their prediction accuracy can be treated as the ground truth to compare the output of the proposed system model. The predicted confidence interval for tumor annotation using neural network and Neurologists were shown in Fig 4 and Fig 5 for Meningioma and Glioma respectively. According to the figures, most of the time average confidence level for tumor annotation obtained using the proposed model tally extremely closely with the neurologists' prediction results. Because of this similarity, we can deduce that the proposed model would work well in the field, to aid the medical professionals with a satisfying level of accuracy.

Further, we noticed a significant difference in the rating level between neurologists and MO/HO as in Fig 6 and Fig 7 for both Glioma and Meningioma identification. This is because the medical image diagnosis is highly dependent on the experience level of the doctors and the level of training they had on subspecialties of medicine, such as neurology, radiology, and neurosurgery. According to fig 6 and fig 7, the prediction from the MO/HO could carry more errors than the neurologists' prediction. Since our proposed algorithm was able to perform accurately in the same capacity as of neurologists, this proposed model can be used to assist the diagnostic process of brain tumors in the absence of expertise, so that the patient can be directed for further medical examination by a qualified specialist.

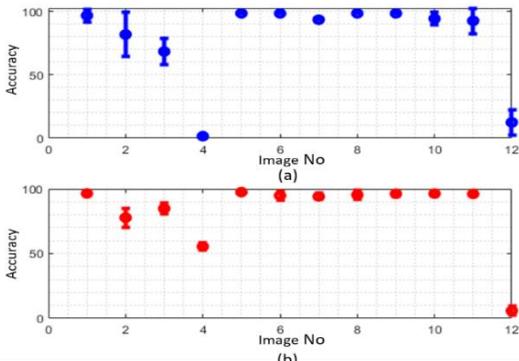

Fig. 4 Subjective evaluation for Meningioma
Between Neurologist (a) and proposed network (b)

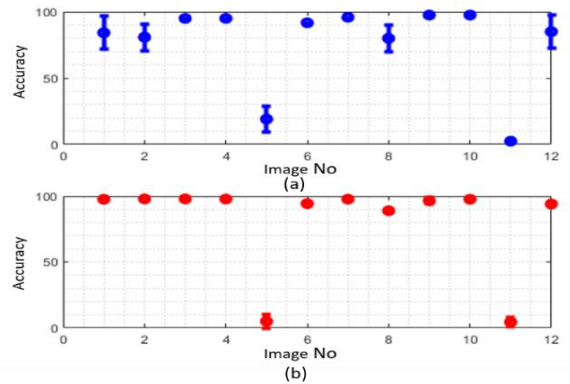

Fig 5 Subjective evaluation for Glioma
Between Neurologist (a) and proposed network (b)

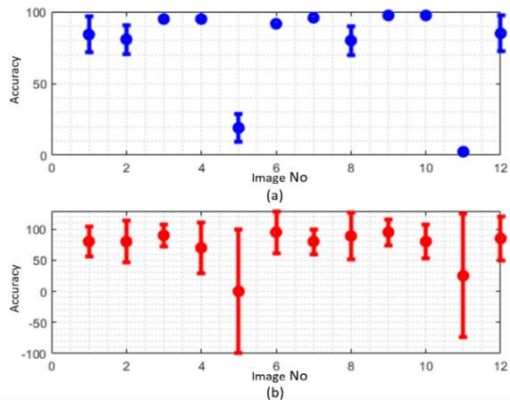

Fig 6 Subjective evaluation for Glioma
Between MO/HO and Neurologist

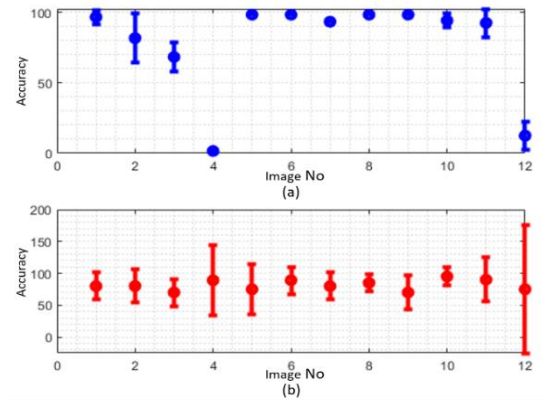

Fig 7 Subjective evaluation for Meningioma
Between MO/HO and Neurologist

# 5 Conclusion

The article presents a feasibility study carried out for deep learning based MRI tumor segmentation. The deep learning algorithm consists of a typical CNN for tumor classification into Meningioma and Glioma, and a faster R-CNN based model for tumor localization of both cases. Authors used 5 fold cross validation with figshare dataset to train the network. For the segmentation task, a Prewitt edge detection algorithm was used and objective quality matrices such as confidence interval, dice score, accuracy and BDE are calculated to evaluate the performance. The system was able to segment a tumor with accuracy of 91.25% , dice score of 0.90 and a confidence interval of 94.78%  Further, we tested the performance of our system both using objective and subjective quality metrics. A subjective quality evaluation was carried out by following DSIS protocol. In there, our model was validated by expertise in the field with a high confidence interval. Hence we can conclude that our model can be used as a reliable aid for brain tumor classification and segmentation in low human resource, expertise, environments.

## Acknowledgment

The authors wish to acknowledge the insightful and extremely helpful reviews and comments provide by the Neurologists, Radiologists, Medical Officers (MO) and Intern Home Officers (HO) in Teaching Hospitals and district hospitals in the country. A special thanks goes to Dr. S.C Weerasinghe, Neurologist, at Teaching hospital Anuradhapura for his guidance during the design and validation stages of the research. Also we thank all the medical professionals who took part in our subjective results analysis stage, despite their busy schedule.